\documentclass[aps, prl, reprint, amsfonts,
  amssymb, amsmath, showpacs, letterpaper]{revtex4-1}

\usepackage{braket, bm}

\usepackage{color,graphicx}

\begin{document}

\title{Radiative Screening of Fifth Forces}

\author{Clare Burrage}
\email{clare.burrage@nottingham.ac.uk}
\affiliation{School of Physics and Astronomy, University of Nottingham, Nottingham NG7 2RD, United Kingdom}

\author{Edmund J. Copeland}
\email{edmund.copeland@nottingham.ac.uk}
\affiliation{School of Physics and Astronomy, University of Nottingham, Nottingham NG7 2RD, United Kingdom}

\author{Peter Millington}
\email{p.millington@nottingham.ac.uk}
\affiliation{School of Physics and Astronomy, University of Nottingham, Nottingham NG7 2RD, United Kingdom}





\begin{abstract}
We describe a symmetron model in which the screening of fifth forces arises at the one-loop level through the Coleman-Weinberg mechanism of spontaneous symmetry breaking. We show that such a theory can avoid current constraints on the existence of fifth forces but still has the potential to give rise to observable deviations from general relativity, which could be seen in cold atom experiments.
\end{abstract}

\maketitle
The mystery of dark energy has motivated much study of scalar-tensor theories~\cite{Copeland:2006wr,Clifton:2011jh}. However, the associated scalar fifth force has not been detected to date, and so either the matter coupling must be fine-tuned or this fifth force must be screened in local environments. This has attracted significant experimental interest, with proposals to test screening models being made across cosmology~\cite{AmendolaLivRev}, astrophysics~\cite{Jain:2012tn}, and the fields of cold atoms~\cite{Hamilton:2015zga,Brax:2016wjk,Burrage:2016rkv} and high-precision optics~\cite{Anastassopoulos:2015yda}. In existing models, this screening arises at the level of the classical action, and one has to worry about radiative stability~\cite{Joyce:2014kja}. In this Letter, we consider a screening mechanism that emerges instead at the one-loop level by virtue of radiative corrections, and we demonstrate that additional loop corrections are sub-leading. Nevertheless, the behaviour of the scalar fifth force is analogous to the symmetron model, first introduced in Refs.~\cite{Hinterbichler:2010es,Hinterbichler:2011ca}.

In the original symmetron model, the scalar fifth force is screened from local tests of gravity as a result of tree-level spontaneous symmetry breaking. This theory has the classical potential 
\begin{equation}
\tilde{V}(\varphi) \equiv V(\varphi)-\mathcal{L}_m[g] = -\frac{1}{2}\mu^2 \varphi^2+\frac{1}{4}\lambda\varphi^4-\mathcal{L}_m[g]\;,
\end{equation}
with the scalar field $\varphi$ coupled universally to matter fields, having Lagrangian density $\mathcal{L}_m$, through the Jordan-frame metric $g_{\mu\nu}$. The latter is related to the Einstein-frame metric $\tilde{g}_{\mu\nu}$ via the conformal transformation $g_{\mu\nu}=A^2(\varphi)\tilde{g}_{\mu\nu}$, where the coupling function $A(\varphi)$~is
\begin{equation}
\label{eq:couplingfunc}
A(\varphi)\ =\ 1\:+\: \frac{\varphi^2}{2M^2}\: +\:\mathcal{O}\left(\frac{\varphi^4}{M^4}\right)\;,
\end{equation}
and the scale $M$ determines the matter coupling. Earlier work studied a similar model but with different motivation~\cite{Pietroni:2005pv,Olive:2007aj}, and string-inspired models, with similar phenomenology, have also been proposed~\cite{Damour:1994zq,Brax:2011ja}.

The classical equation of motion for the symmetron~is
\begin{equation}
\Box\,\varphi\ =\ \frac{\mathrm{d} V}{\mathrm{d}\varphi}\: +\:\tilde{\mathcal{T}}\,\frac{\mathrm{d} A}{\mathrm{d}\varphi}\;,
\end{equation}
where $\tilde{\mathcal{T}}$ is the trace of the Einstein-frame energy-momentum tensor of the local matter fields. When this matter is static and non-relativistic, we can treat it as a pressureless perfect fluid.  In this case, the classical Einstein-frame potential of the symmetron becomes
\begin{equation}
\tilde{V}(\varphi)\ =\ \frac{1}{2}\,\left(\frac{\rho}{M^2}\:-\:\mu^2\right)\varphi^2\: +\:\frac{1}{4}\,\lambda\,\varphi^4\;,
\end{equation}
where $\rho$ is the local matter energy density. Whether the coefficient of the quadratic term is positive or not and, as a result, whether the $\mathbb{Z}_2$ symmetry ($\varphi \rightarrow -\,\varphi$) is spontaneously broken or not depends on the relative values of $\rho/M^2$ and $\mu^2$. Thus, taking $\mu^2>0$ and $\lambda>0$, the symmetry is spontaneously broken in regions of low density and restored when the local density is high enough.

On a test particle of unit mass, the symmetron field mediates a fifth force
\begin{equation}
\vec{F}_{\rm sym}\ =\ \vec{\nabla} A(\varphi)\ =\ \frac{\varphi}{M^2}\,\vec{\nabla}\varphi\;.
\end{equation}
Thus, if the universe is always sufficiently dense that the $\mathbb{Z}_2$ symmetry is everywhere restored, we have $\varphi=0$, and the classical symmetron-mediated force vanishes. Instead, if the universe is in the symmetry-broken phase today, dense concentrations of matter can be enough to restore the symmetry locally.

Inside a spherically symmetric source of radius $R$ and density $\rho_{\mathrm{in}} >  \mu^2 M^2$, the classical potential can be approximated around the minimum at $\varphi=0$ as
\begin{equation}
\tilde{V}(\varphi)\big|_{\varphi\,\sim\,0}\ \approx\ \frac{1}{2}\,m_{\rm in}^2\,\varphi^2\;,
\end{equation}
where $m_{\rm in}^2=\rho_{\rm in}/M^2-\mu^2>0$. Outside the source, where the background density is $\rho_{\rm out}<\mu^2M^2$, the classical~potential can be approximated around the true minima as
\begin{equation}
\tilde{V}(\varphi)\big|_{\varphi\,\sim\,\pm v}\ \approx \ \frac{1}{2}\,m_{\rm out}^2(\varphi\mp v)^2\;,
\end{equation}
where
\begin{equation}
\label{eq:treevev}
v \ \equiv\ m_{\rm out}/\sqrt{\lambda}\;,
\end{equation}
$m_{\rm out}^2= 2(\mu^2-\rho_{\rm out}/M^2)>0$ and we have neglected a constant shift in the potential.

In Ref.~\cite{Hinterbichler:2011ca}, the symmetry-breaking scale is chosen close to the cosmological density today, i.e.~$\mu^2 M^2\sim H_0^2M_{\rm Pl}^2$, where $H_0$ is the present-day Hubble scale, and the symmetron force in vacuum is required to have approximately gravitational strength, i.e.~$v/M^2 \sim 1/M_{\rm Pl}$. Here, $M_{\rm Pl}\equiv (8\pi G)^{-1/2}$ is the reduced Planck mass, where $G$ is Newton's gravitational constant. Assuming $m_{\rm out}r\ll 1$, we can find the general form of the symmetron field around the source:
\begin{equation}
\label{eq:profile}
\varphi(r) = \frac{\pm\,v}{m_{\rm in}r}\left\{\!\!\begin{array}{lc}
\frac{\sinh m_{\rm in} r}{\cosh m_{\rm in}R}\;, \ 0 < r < R\\[0.5em]
\bigg[\frac{\sinh m_{\rm in} R}{\cosh m_{\rm in}R}+m_{\rm in}(r-R)\bigg]\,, \ R < r\;.
\end{array}\right.
\end{equation}
When the size of the source is much bigger than the Compton wavelength of the symmetron field in its interior, i.e.~$m_{\rm in}R \gg 1$, symmetry is restored as $r\to 0$, and we are in the \emph{screened regime}. For $r\gg R$, the symmetron-mediated force is then given by
\begin{equation}
\frac{F_{\rm sym}}{F_N}\ =\  \frac{6 v^2}{\rho_{\rm in} R^2}\left(\frac{M_{\rm Pl}}{M}\right)^2\left(1-\frac{R}{r}\right)\ \ll\ 1\;,
\end{equation}
where $F_N$ is the Newtonian gravitational force. On the other hand, if $m_{\rm in}R \ll 1$, we do not reach the symmetry restored phase as $r\to 0$ and are instead in the \emph{unscreened regime}, and (for $r\gg R$)
\begin{equation}
\frac{F_{\rm sym}}{F_N}\ =\  \frac{2 v^2}{M^2}\left(\frac{M_{\rm Pl}}{M}\right)^2\ \approx\ 2\;.
\end{equation}
The symmetron force between test particles in vacuum can have gravitational strength whilst still evading current bounds from observations on solar-system scales so long as the matter coupling $M \lesssim 10^{-4} M_{\rm Pl}$~\cite{Hinterbichler:2011ca,Amol}.

The symmetron model described above exhibits symmetry breaking at tree level in regions of low matter density. We will now consider a symmetron model in which the symmetry breaking arises \emph{radiatively} in regions of low matter density via the Coleman-Weinberg mechanism~\cite{Coleman:1973jx}. We begin with the following classical action~\cite{Signs}:
\begin{equation}
S\ =\ \int\mathrm{d}^4x\,\sqrt{-\,g}\;\bigg[\frac{1}{2}\,F(\phi) \mathcal{R}\:-\:\Lambda\:+\mathcal{L}\:+\:\mathcal{L}_m\bigg]\;,
\end{equation}
where $\mathcal{R}$ is the Ricci scalar, $\Lambda$ is the bare cosmological constant, which we hereafter neglect, and we work in units of the reduced Planck mass (i.e.~$M_{\rm Pl}=1$) unless otherwise stated. In order to remain in the regime of validity of the Coleman-Weinberg mechanism, the symmetry-breaking vacua for the Brans-Dicke-type scalar field $\phi(x)\equiv\phi_x$ are induced through a coupling to a massless scalar field $X(x)\equiv X_x$:
\begin{align}
-\,\mathcal{L} = \frac{1}{2}\,\phi_{,\mu}\phi^{,\mu}+\frac{1}{2}\,X_{,\mu}X^{,\mu}+\frac{\lambda}{4}\,\phi^2\,X^2+\frac{\kappa}{4!}\,X^4\;.
\end{align}
where $\lambda,\kappa>0$. We employ the signature convention $(-,+,+,+)$. For technical simplicity in what follows, we have set to zero a quartic self-interaction for the field $\phi$. Finally, $\mathcal{L}_m$ is the matter Lagrangian, and we take a non-minimal coupling of the form
\begin{equation}
F(\phi)\ =\ 1\:+\:\frac{\phi^2}{M^2}\;,
\end{equation}
motivated by Eq.~\eqref{eq:couplingfunc}.

We choose to work in the Jordan frame within an effective field theory (EFT) framework, neglecting the direct couplings to the Standard Model (SM) degrees of freedom that are generated via graviton exchange. These couplings appear in the Einstein frame after the Weyl transformation of the matter action and are suppressed by at least the ratio of the electroweak scale (which we take to be of the order of the Higgs vev $v_h=246\ {\rm GeV}$) to the scale $M$. In spite of the absence of explicit couplings to matter fields in the Jordan frame, the geodesic equation still contains terms that can be interpreted as a scalar fifth force, reflecting the classical equivalence of the Einstein and Jordan frames. Moreover, in the small-field regime, $\varphi/M \ll 1$ ($\varphi\equiv\braket{\phi}$), the canonically-normalized Einstein-frame field $\tilde{\varphi}$ is equal to the Jordan-frame field $\varphi$ at leading order:
\begin{equation}
\tilde{\varphi}\: =\: \int\limits_0^{\tilde{\varphi}}\!{\rm d}\varphi\,\Bigg[\frac{F(\varphi)\:+\:\frac{3}{2}\,F'{}^2(\varphi)}{F^2(\varphi)}\Bigg]^{\frac{1}{2}}\: =\: \varphi\bigg[1\:+\:\mathcal{O}\bigg(\frac{\varphi^2}{M^2}\bigg)\bigg]\;.
\end{equation}

Working in the Jordan frame has the advantage that we can keep physical scales distinct and more clearly identify our approximations. It should be stressed, however, that strictly identical results would be obtained in the Einstein frame at the same level of approximation. The EFT treatment remains predictive so long as $v/M< 1$ and the couplings of the scalar sector $\lambda,\kappa> v_H^2/M^2$.

In order to derive the one-loop effective potential, we make the following simplifying approximations:
\begin{itemize}
\item [(i)] The gravitational sector is treated as a classical source, i.e.~we neglect classical and quantum gravitational perturbations.

\item [(ii)] We assume a Minkowski space-time background with constant field configurations $\varphi\equiv\braket{\phi}$ and $\chi\equiv\braket{X}$ when performing the loop integrals.
\end{itemize}
As such, we neglect non-renormalizable operators generated by gravitational interactions, which is appropriate within the EFT description, and the effect of field gradients, which is negligible so long as the size of the source is not comparable to the Compton wavelength of the symmetron.

We require the functional Hessian matrix of the scalar sector of the action, whose elements are:
\begin{gather}
\Delta^{-1}_{\phi\phi}(x,y)\: \equiv\: \frac{\delta^2 S}{\delta \phi_x\delta\phi_y}\bigg|_{\substack{\phi=\varphi\\ X=\chi}}\: = \: \delta^{4}_{xy}\big(\Box-m_{\varphi}^2\big)\;,\nonumber\\
\Delta^{-1}_{\phi X}(x,y)\: \equiv\: \frac{\delta^2 S}{\delta \phi_x\delta X_y}\bigg|_{\substack{\phi=\varphi\\ X=\chi}}\: = \: \delta^{4}_{xy}\big(-\lambda\varphi\chi\big)\;,\nonumber\\
\Delta^{-1}_{XX}(x,y)\: \equiv\: \frac{\delta^2 S}{\delta X_x\delta X_y}\bigg|_{\substack{\phi=\varphi\\ X=\chi}}\: = \: \delta^{4}_{xy}\big(\Box-m_{\chi}^2\big)\;,
\label{eq:Hess}
\end{gather}
where
\begin{gather}
m_{\varphi}^2\ =\ m_{\mathcal{T}}^2\:+\:\frac{\lambda}{2}\,\chi^2\;,\quad m_{\mathcal{T}}^2\ =\ -\,\frac{1}{2}\,\mathcal{R}\,\partial_{\varphi}^2F(\varphi)\;,\nonumber\\
m_{\chi}^2\ =\ \frac{\lambda}{2}\,\varphi^2\:+\:\frac{\kappa}{2}\,\chi^2\;,
\end{gather}
and $\delta^4_{xy}\equiv\delta^4(x-y)$ is the Dirac delta function.

In order to find the explicit form of the background-dependent mass $m_{\mathcal{T}}^2$, we make use of the Jordan-frame Einstein equations, which take the form
\begin{equation}
F(\varphi)G_{\mu\nu}\ =\ F_{;\mu\nu}(\varphi)\:-\:g_{\mu\nu}\Box F(\varphi)\:+\:\mathcal{T}_{\!\mu\nu}\;,
\end{equation}
where $G_{\mu\nu}=\mathcal{R}_{\mu\nu}-\frac{1}{2}g_{\mu\nu}\mathcal{R}$ is the Einstein tensor, and $\mathcal{T}_{\!\mu\nu}$ is the energy-momentum tensor of the scalar and matter sectors. From the trace of the Einstein equations, we find (for the constant background field configurations)
\begin{equation}
\label{eq:trace}
-\,F(\varphi)\mathcal{R}\ = \ \mathcal{T}\;,
\end{equation}
giving
\begin{equation}
\label{eq:mT}
m_{\mathcal{T}}^2\  =\ \frac{\mathcal{T}}{F(\varphi)}\,\frac{\partial F(\varphi)}{\partial \varphi^2}\;.
\end{equation}
Neglecting the contribution to the trace of the energy-momentum tensor from the scalar sector and treating the matter degrees of freedom as a pressureless perfect fluid ($\mathcal{L}_m=\rho$), we have $\mathcal{T}=\rho$. For $\varphi/M\ll 1$, $F(\varphi)\sim 1$, and the background-dependent mass is given by
\begin{equation}
m_{\mathcal{T}}^2\ \simeq\ \rho/M^2\;.
\end{equation}
Thus, in vacuum, $m_{\mathcal{T}}^2=0$, and we have a classically scale invariant theory, whose one-loop corrections suffer logarithmic infra-red divergences. In order to regularize these divergences, we introduce a mass scale $m$, which is, via dimensional transmutation, translated to a symmetry-breaking scale $v$ by the Coleman-Weinberg mechanism~\cite{Coleman:1973jx}.

The one-loop contribution to the effective potential~\cite{Jackiw:1974cv} ($\hbar=1$) is given by
\begin{equation}
\label{eq:oneloopform}
V^{(1)}(\varphi)\ =\ \frac{i}{2\,\mathcal{V}}\,\mathrm{Tr}\,\ln\,\mathrm{det}\,\bm{\Delta}^{-1}\:+\:\delta V\;,
\end{equation}
where $\mathcal{V}$ is a four-volume factor, the determinant runs over the elements of the functional Hessian matrix in Eq.~\eqref{eq:Hess}, and $\delta V$ contains the counterterms. These take the general form
\begin{align}
&\delta V\ =\ \delta \Lambda\:+\:\frac{1}{2}\,\delta F\,\mathcal{R}\:+\:\delta\mathcal{L}\;,
\end{align}
where $\delta \Lambda\sim \Lambda_{\rm UV}^4$ and $\delta F\sim \Lambda_{\rm UV}^2$ are constant functions of $\varphi$, and $\delta \mathcal{L}$ contains the counterterms of the scalar sector~\cite{Counters}. We choose to fix the latter by the following renormalization conditions, which leave the mass and couplings unchanged at the renormalization points:
\begin{gather}
\frac{\partial^4 V}{\partial \varphi^4}\bigg|_{\substack{\varphi=0\\\chi=m}} = 0\;,\quad
\frac{\partial^4 V}{\partial \varphi^2\partial \chi^2}\bigg|_{\substack{\varphi=0\\ \chi=m}}\ =\ \lambda\;,\quad \frac{\partial^4 V}{\partial \chi^4}\bigg|_{\substack{\varphi=0\\\chi=m}} = \kappa\;,\nonumber\\
\frac{\partial^2 V}{\partial \varphi^2}\bigg|_{\varphi,\chi=0} = m_{\mathcal{T}}^2\;,\quad
\frac{\partial^2 V}{\partial \chi^2}\bigg|_{\varphi,\chi=0} = 0\;.
\end{gather}

Given the approximations listed earlier, the trace in Eq.~\eqref{eq:oneloopform} can be performed conveniently in momentum space by first Wick rotating to Euclidean space and then introducing the ultra-violet (UV) cut-off $\Lambda_{\rm UV}$ on the three-momentum integral. One then finds that the global minima of this one-loop effective potential lie along the line $\chi=0$ (see~Ref.~\cite{Garbrecht:2015yza}, where an $O(N)$-symmetric generalization of this model was analyzed in the context of vacuum decay). Hereafter setting $\chi=0$, the mass matrix, whose elements appear in Eq.~\eqref{eq:Hess}, has eigenvalues
\begin{equation}
m_{\varphi}^2\ =\ m_{\mathcal{T}}^2\;,\qquad m_{\chi}^2\ =\ \frac{\lambda}{2}\,\varphi^2\;.
\end{equation}
The contribution from the first eigenvalue yields one-loop corrections that are a function of $\mathcal{R}$. However, these terms do not carry any explicit dependence on $\varphi$. Since we are interested only in contributions that have such a dependence, these terms may be neglected along with contributions to the cosmological constant. The relevant renormalized one-loop terms are then
\begin{equation}
V^{(1)}(\varphi)\ = \ \bigg(\frac{\lambda}{16\pi}\bigg)^2\,\varphi^4\,\bigg(\!\ln\frac{\varphi^2}{m^2}\:-\:Y\bigg)\;,
\end{equation}
where
\begin{align}
Y\ &=\ \frac{1}{(1-y)^3}\bigg[4\big[3-y\big(2y+13\big)\big]\nonumber\\&\qquad+\:\big(3+y\big)\big[3+y\big(6-y)\big]\bigg(\!\ln\,y+\frac{3}{2}\bigg)\bigg]\;,
\end{align}
and $y\equiv \kappa/\lambda$ is the ratio of the couplings. Having used an auxiliary field to induce the symmetry breaking, we obtain dependence on the ratio of the couplings only, with the exception of an overall scaling of the one-loop term. Hence, so long as $\kappa\sim\lambda$, we remain always in the region of validity of the one-loop approximation. In addition, within the regime of validity of the EFT, matter loops only contribute corrections to $M_{\rm Pl}$ and the cosmological constant $\Lambda$. As such, this mechanism can be regarded as radiatively stable in the sense that the one-loop results presented here are predictive. 

Taking $\kappa\to \lambda$, $Y=17/6$, and the relevant part of the renormalized one-loop effective potential simplifies to
\begin{equation}
\label{eq:Veff}
V(\varphi)\ = \ \frac{1}{2}F(\varphi)\mathcal{R}\:+\:\bigg(\frac{\lambda}{16\pi}\bigg)^2\,\varphi^4\,\bigg(\!\ln\frac{\varphi^2}{m^2}\:-\:\frac{17}{6}\bigg)\;.
\end{equation}
The partial derivative of this potential with respect to $\varphi$ is given by~\cite{VariationOrder}
\begin{equation}
\label{eq:derivativepot}
V'(\varphi)\ =\ \ m_{\mathcal{T}}^2\,\varphi\:+\:\bigg(\frac{\lambda}{8\pi}\bigg)^2\,\varphi^3\,\bigg(\!\ln\frac{\varphi^2}{m^2}\:-\:\frac{7}{3}\bigg)\;.
\end{equation}
Equation \eqref{eq:derivativepot} has five roots: we find an extremum at $\varphi=0$, two minima at
\begin{equation}
\label{eq:min}
\varphi\ =\ \pm\,v_{\rm min}(z)\ \equiv\ \pm\,m\,e^{7/6}\,\bigg(\frac{z}{W_0(z)}\bigg)^{1/2}\;,
\end{equation}
where $W_0$ is the principal branch of the Lambert $W$ and
\begin{equation}
z\ \equiv\ -\,e^{-7/3}\,\bigg(\frac{8\pi}{\lambda}\,\frac{m_{\mathcal{T}}}{m}\bigg)^2\;,
\end{equation}
and two maxima at
\begin{equation}
\label{eq:max}
\varphi\ =\ \pm\,v_{\rm max}(z)\ \equiv\ \pm\,m\,e^{7/6}\,\bigg(\frac{z}{W_{-1}(z)}\bigg)^{1/2}\;,
\end{equation}
where $W_{-1}$ is the lower real branch of the Lambert $W$.

In the limit $m_{\mathcal{T}}\to 0$, we have two symmetry-breaking minima at
\begin{equation}
\varphi \ =\  \pm\,v\ \equiv\ \pm\,m\,e^{7/6}
\end{equation}
and a ``flat maximum'' at the origin. Around the minima, the potential is approximately
\begin{equation}
V(\varphi)\big|_{\varphi\,\sim\,\pm\,v_{\rm min}(z)}\ \approx\ \frac{1}{2}\,m_{\rm min}^2(z)(\varphi-v_{\rm min}(z))^2\;,
\end{equation}
where
\begin{equation}
m_{\rm  min}^2(z)\ =\ -\,2\,m_{\mathcal{T}}^2\bigg(1+\frac{1}{W_0(z)}\bigg)\;.
\end{equation}
Hence, in the cosmological vacuum today, we find $m_{\rm min}^2\approx \lambda^2v^2/32/\pi^2$, corresponding to a Compton wavelength
\begin{equation}
\label{eq:Compton}
\bigg(\frac{l_{\rm Comp}}{\rm cm}\bigg)\ \simeq\ \frac{10^{-30}}{\lambda}\,\bigg(\frac{M_{\rm{Pl}}}{v}\bigg)\;.
\end{equation}

When $m_{\mathcal{T}}$ is large, we have one minimum at $\varphi =0$, and the symmetry is restored. This occurs at the branch point of the Lambert $W$ when $z=-\,e^{-1}$. Thus, symmetry is restored when $m_{\mathcal{T}} > \lambda\,v/8/\pi$ or, equivalently,
\begin{equation}
\label{eq:restorationcondition}
\rho \ >\ \bigg(\frac{\lambda}{8\pi}\bigg)^2\,e^{4/3}\,m^2\,M^2\;.
\end{equation}
The field $\phi$ acts as a symmetron, the behaviour of which is determined radiatively.

\begin{figure}
\centering
\includegraphics[scale=0.47]{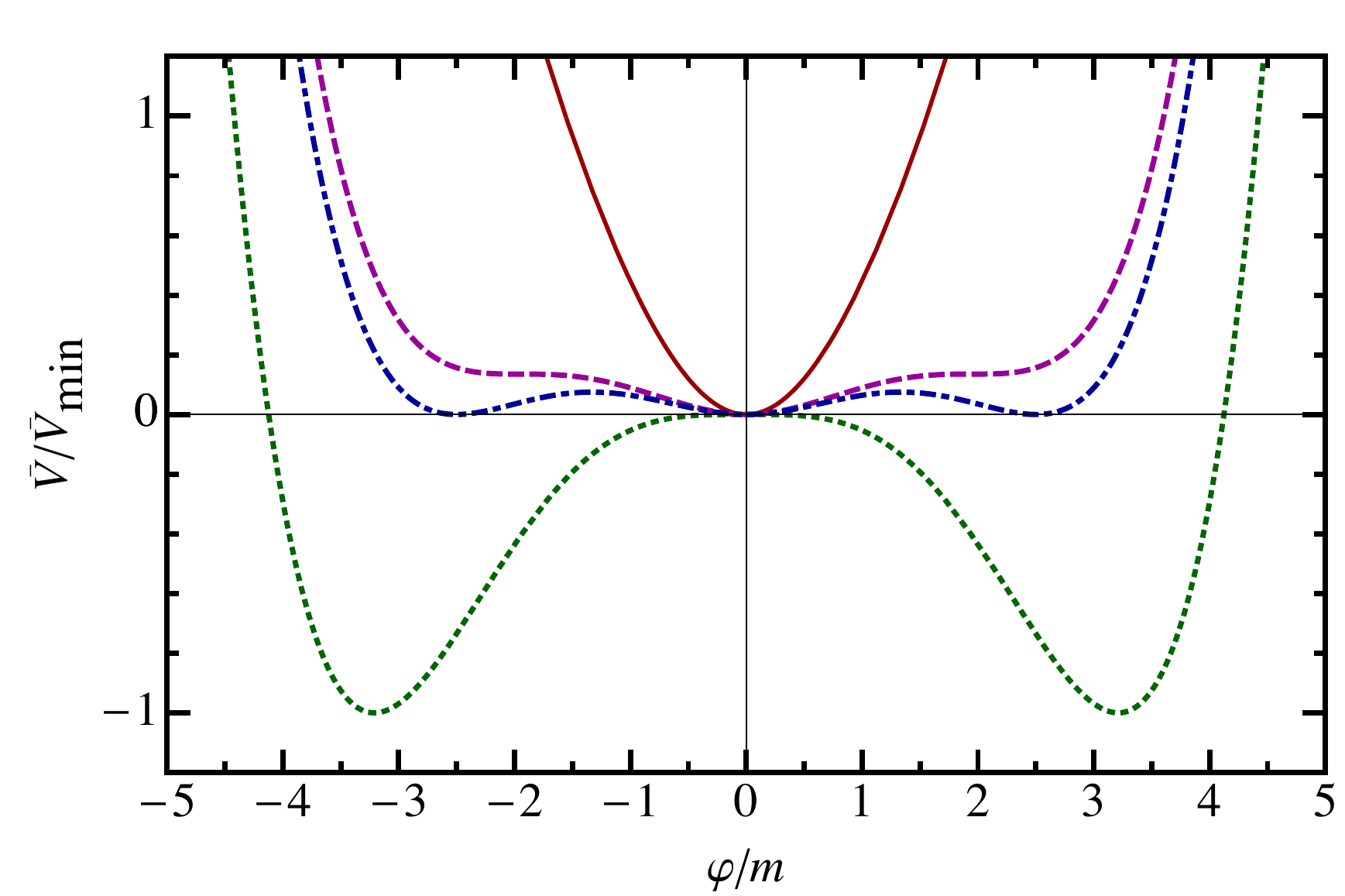}
\caption{\label{fig:potential} Plot of the shifted one-loop potential $\bar{V}(\varphi)$, normalized to its minimum value, as a function of $\varphi/m$ in the symmetry-broken phase (dotted green) for $m_{\mathcal{T}}\to 0$, at the degenerate point (dash-dotted blue), at the critical point (dashed magenta) and in the symmetric phase (solid red).}
\end{figure}

In order to illustrate this behaviour, we define a shifted potential $\bar{V}(\varphi)$ by integrating Eq.~\eqref{eq:derivativepot} with respect to $\varphi$ subject to the condition $\bar{V}(0)=0$. This is shown in Fig.~\ref{fig:potential} for the symmetry-broken and symmetry-restored phases, as well as at the ``critical point'', where the minima and maxima given by Eqs.~\eqref{eq:min} and \eqref{eq:max} merge into inflection points. Figure~\ref{fig:potential} also shows the form of the potential at the ``degenerate point''
\begin{equation}
\label{eq:degenerate}
\rho\ =\ \frac{1}{2}\bigg(\frac{\lambda}{8\pi}\bigg)^2e^{11/6}\,m^2\,M^2\;,
\end{equation}
at which there are three degenerate minima. Below the critical point, the presence of the potential barrier between local and global minima allows for \emph{density-driven} first-order phase transitions in the low-temperature regime.

In the high-temperature regime, thermal corrections dominate, and we must replace Eq.~\eqref{eq:Veff} by the thermal effective potential. Its high-temperature expansion is~\cite{Carrington:1991hz}
\begin{align}
V(\varphi)\ &= \ \frac{\lambda T^2}{48}\,\varphi^2\:-\:\frac{\lambda^{3/2}T}{12\pi}\bigg(\frac{\varphi^2}{2}+\frac{T^2}{12}\bigg)^{\!3/2}\nonumber\\&\qquad +\:\bigg(\frac{\lambda}{16\pi}\bigg)^{\!2}\varphi^4\bigg(\ln\frac{32\pi^2T^2}{\lambda\,m^2}\:-\:\frac{17}{6}\bigg)\;,
\end{align}
where  $T$ is the temperature. This potential exhibits a first-order thermal phase transition~\cite{Linde:1980tt} with a critical temperature
\begin{equation}
T_c\ \simeq\ \frac{e^{11/4}}{4\sqrt{2}\pi}\,\lambda^{1/2}\,m\ \sim\ \frac{1}{4}\,\lambda^{1/2}\,v\;.
\end{equation}
Moreover, the ratio $v_c/T_c\sim \lambda^{-1/2}> 1$ for $\lambda <1$, where $v_c$ is the value of the field in the critical minimum, signifying that the phase transition is strongly first order, having the potential to produce relic gravitational waves~\cite{Witten:1984rs, Kosowsky:1991ua,Caprini:2009fx}. An analogous calculation for the original symmetron model yields a critical temperature $T_c\simeq\sqrt{2}\,v$, parametrically larger than that of the present model (for small couplings). In addition, the original symmetron model, having $v_c/T_c\sim \lambda^{1/4}< 1$, can yield a strong first-order phase transition only if matter loops can deliver a sufficiently large cubic self-interaction.

\begin{figure}
\includegraphics[scale=0.47]{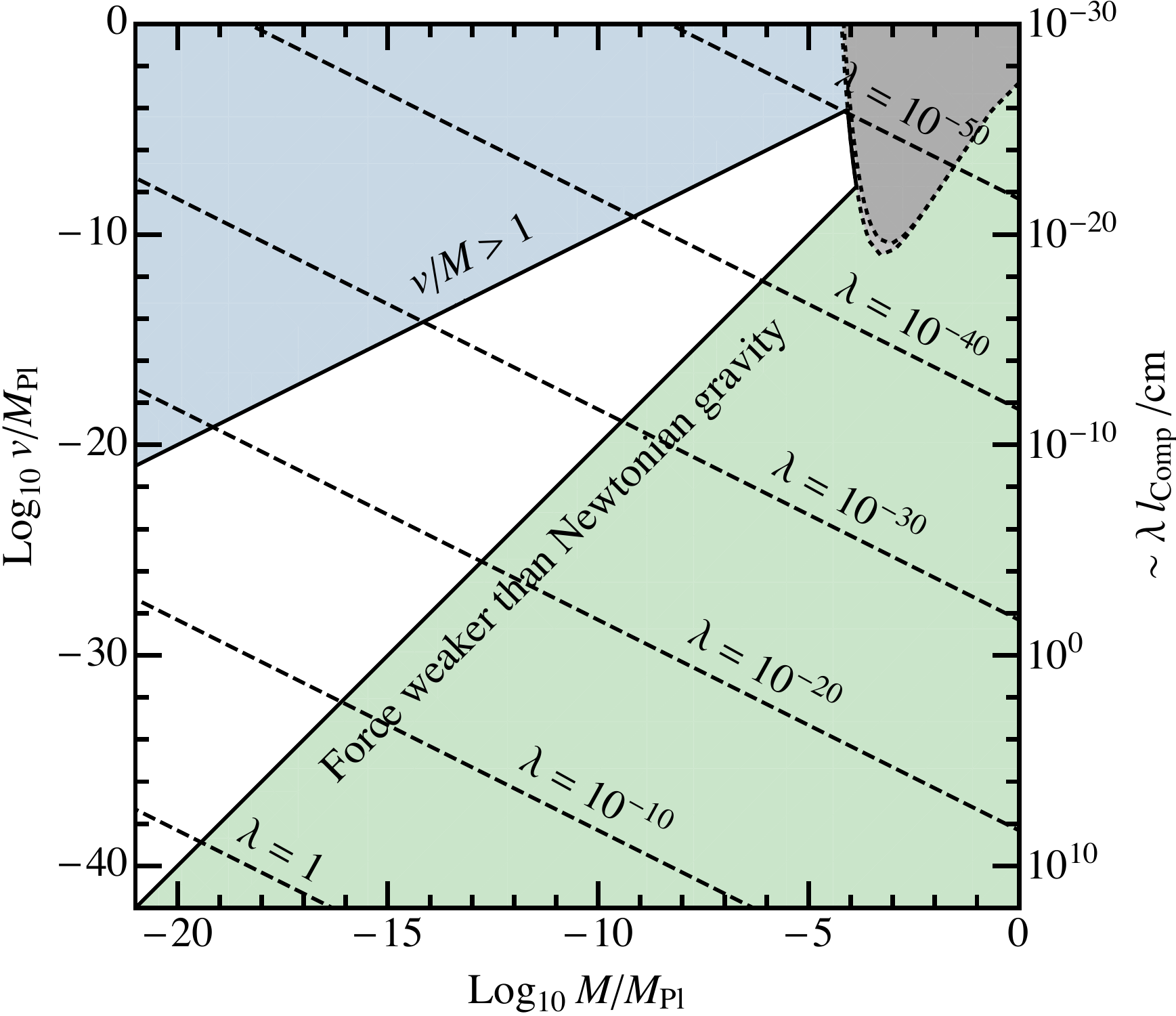}
\caption{\label{fig:constraints} Constraints on the scales $v$ and $M$. The upper (blue) region, $v/M>1$ [Eq.~\eqref{eq:EFTconstraint}], lies outside the validity of the EFT. In the lower (green) region the fifth force is weaker than Newtonian gravity $(\alpha<1)$ [cf.~Eq.~\eqref{eq:alpha}]. The overlapping grey regions in the top right are excluded by constraints on PPN parameters [Eq.~\eqref{eq:PPNconstraint}]; dark and light grey correspond to $m_{\rm out}R_*\ll1$ and $m_{\rm out}R_*\gg X$, respectively. For a given value of $\lambda$, the cosmological vacuum is in the symmetry-broken phase today over the region of the $v$-$M$ plane above the corresponding dashed line [Eq.~\eqref{eq:vacuumconstraint}]. The right-hand axis gives $\lambda$ times the Compton wavelength in the cosmological vacuum [Eq.~\eqref{eq:Compton}].}
\end{figure}

Having chosen $\kappa=\lambda$, the model has three free parameters: the coupling $\lambda$, the symmetry-breaking scale $v$ and the coupling scale $M$. These parameters can be further constrained:
\begin{itemize}
\item [(i)] Since $\varphi\in[-\,v,v]$ and assuming a SM matter sector, predictivity of the EFT requires
\begin{equation}
\label{eq:EFTconstraint}
\frac{v}{M_{\rm Pl}}\ <\ \frac{M}{M_{\rm Pl}}\;,\qquad 
\lambda \ >\ \bigg(\frac{v_{H}}{M}\bigg)^{\!2}\;.
\end{equation}
\item [(ii)] We may parametrize the strength of the fifth force relative to Newtonian gravity (for $r\gg R$) by
\begin{equation}
\label{eq:alpha}
\alpha\ \equiv\ \frac{v}{M}\,\frac{M_{\rm Pl}}{M}\;.
\end{equation}
Following Ref.~\cite{Hinterbichler:2011ca}, constraints on parametrized post-Newtonian (PPN) parameters from lunar laser ranging and time-delay experiments made by the Cassini spacecraft then require
\begin{align}
\label{eq:PPNconstraint}
10^{-6}\ &\gtrsim \ \frac{\alpha}{\sqrt{3}}\,\mathrm{max}\bigg(1,2\sqrt{5}\frac{M}{M_{\rm Pl}}\bigg)\,\sinh\bigg(X\frac{R_s}{R_*}\bigg)\nonumber\\&\qquad\times\begin{cases}{\rm sech}\,X\;, & m_{\rm out}R_*\ \ll\ 1\;,\\
X{\rm csch}\,X \;, & m_{\rm out}R_*\ \gg\ X\;.
\end{cases}	
\end{align}
where $X\equiv\sqrt{6\Phi_*}\,M_{\rm Pl}/M$, $\Phi_*\simeq 10^{-6}$ and $R_*\sim 100\ {\rm kpc}$ are the gravitational potential and radius of the Milky Way, and $R_s\sim 10\ {\rm kpc}$ is our distance from the Galactic centre. We note that $\phi$-mediated effective interactions between the field $X$ and matter fields $\psi$, i.e.~$X^2\bar{\psi}\psi$, are suppressed by $\lambda v^2/M^2$.
\item [(iii)] In order to be in the symmetry-broken phase today, the cosmological density ($\rho=3H_0^2M_{\rm Pl}^2$) must be below the degenerate point in Eq.~\eqref{eq:degenerate}:
\begin{equation}
\label{eq:vacuumconstraint}
\bigg(\frac{H_0}{M_{\rm Pl}}\bigg)^{\!2}\ <\ \frac{1}{6}\bigg(\frac{\lambda}{8\pi}\bigg)^2\,e^{-1/2}\,\bigg(\frac{v}{M_{\rm Pl}}\bigg)^{\!2}\,\bigg(\frac{M}{M_{\rm Pl}}\bigg)^{\!2}\;.
\end{equation}
\end{itemize}
These constraints are illustrated in Fig.~\ref{fig:constraints}. By virtue of (i), the maximum Compton wavelength for which this analysis remains predictive is tied to the electroweak scale (or, more generally, the scale of new non-gravitational physics). Saturating the constraints, we find
\begin{equation}
\frac{l_{\rm Comp}}{\rm cm}\ <\ \frac{100}{\alpha}\;,
\end{equation}
giving the generic prediction $l_{\rm Comp}\lesssim 1\ {\rm m}$ for $\alpha\sim 1$. We remark that it would be of interest to include bare portal-type interactions with the SM Higgs field of the form $g\phi^2H^{\dag}H/2$ (see e.g.~Refs.~\cite{Hempfling:1996ht,Chang:2007ki,Englert:2013gz}), as well Yukawa interactions with SM fermions. By tuning these bare couplings against those generated via graviton exchange (and neglected in this analysis), it may be possible to relax the lower bound on the coupling, increasing the maximum achievable Compton wavelength.

Taking $\lambda\sim 10^{-18}$, $v \sim 10^3\ {\rm TeV}$ and $M\sim 10^{-5}M_{\rm {Pl}}$, we can achieve $\alpha\sim 1/100$, $l_{\rm Comp}\sim 1\ {\rm cm}$ and a strong first-order phase transition with $T_c\sim 1\ {\rm MeV}$. This is particularly interesting, as a range of tabletop experiments are currently searching for screened fifth forces over $\sim{\rm cm}$ distance scales~\cite{Brax:2010xx,Brax:2011hb,Upadhye:2012qu,Jenke:2014yel,Hamilton:2015zga,Lemmel:2015kwa,Li:2016tux,Rider:2016xaq} and pulsar timing arrays may be sensitive to the stochastic background of nHz gravitational waves from first-order phase transitions with critical temperatures in the $\rm {MeV}$ range~\cite{Caprini:2010xv}. For shorter Compton wavelengths, we can also obtain phase transitions at around the electroweak scale, potentially having gravitational-wave signatures in the mHz range of the forthcoming LISA satellite array~\cite{Caprini:2015zlo}.

By means of this simple toy model, we have illustrated the feasibility of generating radiatively stable screening mechanisms entirely through quantum corrections. Having shown the phenomenological viability of this model, it would be of interest to study its potential embeddings within the Standard Model and its further implications for both cold atom fifth-force experiments and gravitational-wave observations.

\begin{acknowledgments}
This work was supported by STFC Grant No.~ST/L000393/1 and a Royal Society University Research Fellowship. P.M. would like to thank Bj\"{o}rn Garbrecht for earlier collaboration on radiative symmetry breaking in the context of metastable vacua. C.B. would like to thank Pedro Ferreira for useful discussions. The authors are grateful to Philippe Brax, Bj\"{o}rn Garbrecht, Kurt Hinterbichler, Joerg Jaeckel, Eugene Lim, Antonio Padilla and David Seery for constructive comments.
\end{acknowledgments}


\begin{thebibliography}{99}
\bibitem{Counters}
  Here, since we do not consider the loop corrections to the energy-momentum tensor, we do not include the $\mathcal{O}(\mathcal{R}^2)$ counterterms that would otherwise be necessary.
\bibitem{VariationOrder}
  Note that we must first vary with respect to $\varphi$, keeping the metric fixed, before substituting the trace of the Einstein equation from Eq.~\eqref{eq:trace}.
\end{thebibliography}
\end{document}